# Fermi surface complexity, effective mass, and conduction band alignment in n-type thermoelectric $Mg_3Sb_{2-x}Bi_x$ from first principles calculations

Jiawei Zhang[*] and Bo Brummerstedt Iversen[*]

*Center for Materials Crystallography, Department of Chemistry and iNANO, Aarhus University, DK-8000 Aarhus, Denmark*

Using first principles calculations, we study the conduction band alignment, effective mass, and Fermi surface complexity factor of n-type $Mg_3Sb_{2-x}Bi_x$ ($x$ = 0, 1, and 2) from the full *ab initio* band structure. We find that with increasing Bi content the K and M band minima moves away from the conduction band minimum $CB_1$ while the singly-degenerate Γ band minimum shifts rapidly downward and approaches the conduction band minimum. However, the favorable six-fold degenerate $CB_1$ band minimum keeps dominating the conduction band minimum and there is no band crossing between the Γ and $CB_1$ band minima. In addition, we show that the connection of the $CB_1$ carrier pockets with the energy level close to the band minimum M can strongly enhance the carrier pocket anisotropy and Fermi surface complexity factor, which is likely the electronic origin for the local maximum in the theoretical power factor. Our calculations also show that the band gap, density of states effective mass, Seebeck coefficient, and Fermi surface complexity factor decrease with increasing the Bi content, which is unfavorable to the electrical transport. In contrast, reducing the conductivity effective mass with increasing the Bi content is beneficial to the electrical transport by improving carrier mobility and weighted mobility as long as the detrimental bipolar effect is insignificant. As a result, in comparison with n-type $Mg_3Sb_2$, n-type $Mg_3SbBi$ shows higher power factors and a much lower optimal carrier concentration for the theoretical power factor at 300 K, which can be easily achieved by the experiment.

## I. INTRODUCTION

The practical application of thermoelectric (TE) devices is critically limited by their low TE efficiencies, which are determined by the TE performance of materials used for making the devices.[1,2] The performance of a TE material is characterized by its dimensionless figure of merit, $zT=\alpha^2\sigma T/\kappa$, where $\alpha$ is the Seebeck coefficient, $\sigma$ is the electrical conductivity, $\kappa$ is the thermal conductivity, and $T$ is the absolute temperature. In order to enhance the $zT$ value, one should either reduce the thermal conductivity $\kappa$ or increase the power factor $\alpha^2\sigma$.

$Mg_3Sb_2$-based compounds recently become one of the most intensively investigated TE materials due to the abundant and low-cost constituent elements. These compounds are found to be predominantly p-type,[3-12] which is typically induced by the energetically stable cation vacancies pinning the chemical potential close to the valence bands.[13-15] It is thereby a great advance that the extraordinary n-type TE properties were recently discovered in n-doped $Mg_3Sb_{2-x}Bi_x$.[14,16-30] Superior n-type TE properties with a high $zT$ larger than 1.5 in Te-doped $Mg_{3+x}Sb_{1.5}Bi_{0.5}$ were reported separately by Tamaki *et al.*[18] and Zhang *et al.*[19], where the experimental data reported by Zhang *et al.*[19] is based on reproducing and improving the original work by Pedersen[17] in 2012. Later, several research efforts by Shuai *et al.*,[20] Mao *et al.*,[21] and Chen *et al.*[22] were focused on tuning the carrier scattering mechanism via codoping the transition metals on the Mg sites in order to improve carrier mobility and power factor at low temperatures. Motivated by the earlier work of Zhang *et al*.,[19] increasing the pressing temperature was also confirmed in several reports[23,31] to be effective in tuning the carrier scattering mechanism at low temperatures, which might be explained by the reduced grain boundary electrical resistance at a higher pressing temperature.[27] Moreover, increasing the Bi content in n-type Te-doped $Mg_3Sb_{2-x}Bi_x$ was shown by Imasato *et al*.[25,29] and Shu *et al*.[30] to significantly enhance the low-temperature TE performance, which is even superior to n-type $Bi_2Te_3$.

---
[*]Authors to whom correspondence should be addressed. Electronic addresses: jiaweizhang@chem.au.dk and bo@chem.au.dk

Theoretically, electronic structure and electrical transport properties of n-type $Mg_3Sb_2$-based compounds have been widely studied in various reports.[19,24,32-39] It is found that the electronic origin of the superior n-type electronic transport in $Mg_3Sb_2$-based materials is the complex Fermi surface dominated by the six conducting electron pockets induced by the conduction band minimum (CBM) at the $CB_1$ point.[18,19,24] However, insight into the complexity of Fermi surface especially the carrier pocket anisotropy and its correlation with the electronic transport remain largely unexplored. Alloying with $Mg_3Bi_2$ is found to result in reducing the conductivity effective mass of $Mg_3Sb_2$, which is conducive to improving carrier mobility and weighted mobility.[25,37] The previous theoretical work typically estimated effective mass through fitting the near-edge conduction band that overlooks the nonparabolic and multiple band behavior.[37] Thus, it is desirable to study the effect of the $Mg_3Bi_2$ alloying on the conductivity effective mass extracted from the full *ab initio* band structure for better understanding of the electronic transport. Moreover, the convergence of the band minima at the $CB_1$ and $\Gamma$ points was proposed to rationalize the optimum Bi composition for the electrical transport in n-type $Mg_3Sb_{2-x}Bi_x$,[25,29] which, however, lacks further confirmation.

In this work, we investigate the Fermi surface complexity, the effective masses, the conduction band alignment, and their correlations with the electronic transport in n-type $Mg_3Sb_{2-x}Bi_x$ from the full *ab initio* band structure calculations. The complexity of Fermi surface characterized by $N_v^*K^*$ is dependent on the effective carrier pocket anisotropy $K^*$ and the effective degeneracy of carrier pockets $N_v^*$. In addition to the high valley degeneracy, the carrier pocket anisotropy is also found responsible for the superior Fermi surface complexity. It is revealed that the Fermi surface complexity can be maximized at the energy level slightly above the band minimum at the M point, which can be attributed to the strongly enhanced carrier pocket anisotropy induced by the six-fold $CB_1$ carrier pockets being connected through the M point. The enhanced Fermi surface complexity with the connection of carrier pockets is likely the electronic origin for the characteristic peak in the theoretical power factor at the energy level close to the band minimum M. Regarding the conduction band alignment, we find that the six-fold degenerate band minimum $CB_1$ remains as the CBM and the band crossing between the band minima $\Gamma$ and $CB_1$ expected in earlier reports[25,29] does not occur in n-type $Mg_3Sb_{2-x}Bi_x$. Moreover, it is found that increasing the $Mg_3Bi_2$ alloying composition reduces the density of states effective mass, Seebeck coefficient, and Fermi surface complexity. However, increasing the Bi content in n-type $Mg_3Sb_{2-x}Bi_x$ indeed results in the lighter conductivity effective mass, which is beneficial to enhancing the power factor as long as the bipolar effect is insignificant. Comparing with n-type $Mg_3Sb_2$, another benefit for the $Mg_3Bi_2$ alloying in n-type $Mg_3Sb_{2-x}Bi_x$ is that the corresponding carrier concentration for the optimal theoretical power factor at 300 K is shifted to a much lower value, which is easier to be achieved by the experiment.

## II. COMPUTATIONAL DETAILS

Structural optimizations in this work were conducted using a $6 \times 6 \times 4$ Monkhorst-Pack $k$ mesh and a Hellmann-Feynman force convergence criterion of $10^{-3}$ eV Å$^{-1}$ with the VASP code[40] based on the projector-augmented wave method.[41] Different methods including the PBE functional[42] and HSE06 functional[43] were used for structural relaxations. By comparing the results based on different relaxation schemes (see Table S1), it is found that the conduction band alignment is somewhat sensitive to the structural parameters. In order to obtain a good agreement with the experiment, in this work the lattice parameters are fixed to the experimental values while the atomic positions were relaxed using the HSE06 functional. Electronic structures were computed using the TB-mBJ potential[44] with both the VASP code (a 12 × 12 × 8 $k$ mesh) and the full-electron Wien2k code[45] (a 36 × 36 × 23 $k$ mesh) for comparison. Electronic structure and Boltzmann transport calculations of $Mg_3Sb_2$ with the Wien2k code are reproduced from Refs. 19 and 24. For $Mg_3SbBi$, the calculations were done in the unit cell by replacing one Sb with Bi. The effect of spin orbit coupling (SOC) on the conduction bands was investigated. All VASP calculations were done until the energy converged at the $10^{-6}$ eV level with a plane-wave energy cutoff of 400 eV. For Wien2k calculations, the plane wave cutoff parameter $R_{MT}K_{max} = 9$ was adopted. Full DFT band structures computed with a very dense 36 × 36 × 24 $k$ mesh in the Wien2k code were used for electrical transport calculations. Electrical transport calculations were done with the BoltzTraP package[46] based on the constant scattering time approximation (CSTA). Hall carrier concentration $n_H$ was estimated by $1/eR_H$, where Hall coefficient $R_H$ was extracted from BoltzTraP. The conductivity effective mass is calculated by[47]

$$m_c^* = \frac{ne^2}{\sigma/\tau}, \tag{1}$$

where $n$ is the carrier density, $e$ is the elementary charge, and $\tau$ is the carrier's scattering time. The density of states (DOS) effective mass $m_d^*$ is calculated by solving the single band formula under the acoustic phonon scattering mechanism with the Seebeck coefficient and $n_H$ from BoltzTraP.

## III. RESULTS AND DISCUSSION

The intrinsic electrical transport property of a crystalline solid is closely related to the degeneracy and effective mass of the electronic bands at the band edges. The Seebeck coefficient is governed by the density of states effective mass $m_d^* = N_v^{2/3} m_s^*$, where $N_v$ and $m_s^*$ represent the valley degeneracy of the electronic bands and the effective mass of a single valley, respectively.[48] It is generally accepted that the weighted mobility $\mu(m_d^*/m_e)^{3/2}$ can be used as a descriptor for characterizing the optimum electrical transport performance.[2,48] The weighted mobility can be written as $N_v/m_c^*$ ($N_v/m_s^*$ for an isotropic band) provided that the carrier scattering is dominated by acoustic phonon scattering or alloy scattering. Clearly, in order to achieve excellent electrical properties a high valley degeneracy as well as a light conductivity effective mass is required. It is well known that increasing the valley degeneracy is very effective in optimizing electronic transport through improving Seebeck coefficient without obviously reducing carrier mobility when the effect of the intervalley scattering is unimportant.[2,48-51] The valley degeneracy can be defined as the number of different carrier pockets (for the same type of carriers) existing at a given energy level. In general, the valley degeneracy $N_v$ of a single band extremum is determined by the number of the symmetry equivalent positions in the first Brillouin zone for the specific $k$ point at which the band extremum occurs. It is thereby clear that achieving a high $N_v$ value needs a high-symmetry Brillouin zone and a low symmetry $k$ point where the band extremum occurs.

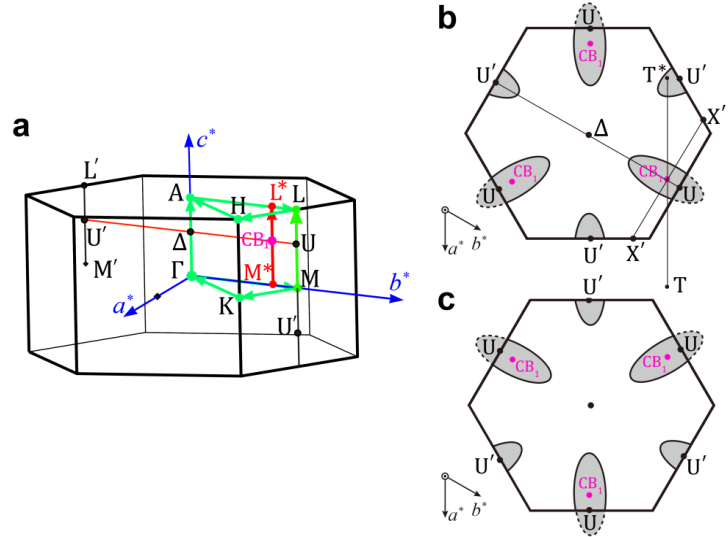

FIG. 1. (Color online) (a) High symmetry $k$ points and $k$-paths in the Brillouin zone of $Mg_3Sb_2$. (b, c) The cross sections containing the $CB_1$ point at the top half (b) and the bottom half (c) of the Brillouin zone. The schematic plot of the six-fold carrier pockets at the $CB_1$ point illustrates that when the carrier pockets expand across the BZ boundary the parts of carrier pockets outside the BZ are moved by one reciprocal lattice vector to the U′ point inside the BZ. The coordinates of the $k$ points in the Figure are Γ (0, 0, 0), K (1/3, 1/3, 0), M (0, 0.5, 0), L (0, 0.5, 0.5), M′ (0, -0.5, 0), L′ (0, -0.5, 0.5), A (0, 0, 0.5), H (1/3, 1/3, 0.5), $M^*$ (0, 0.417, 0), $L^*$ (0, 0.417, 0.5), U (0, 0.5, 0.333), U′ (0, -0.5, 0.333), Δ (0, 0, 0.333), $CB_1$ (0, 0.417, 0.333), $T^*$ (-0.5, 0.417, 0.333), and T (0.5, 0.417, 0.333).

Importantly, understanding electronic transport requires the identification of the near-edge band extrema. Strictly speaking, the band extrema can be defined as the points satisfying $dE/dk = 0$ with the same sign of $d^2E/dk^2$ along different $k$ directions. The band extrema for the same type of carriers are beneficial to the electronic transport as they will form the isolated carrier pockets, which will effectively increase the valley degeneracy. In addition to the actual band extrema, there are pseudo-extremum points for the electronic bands that satisfy $dE/dk = 0$ with the same sign of $d^2E/dk^2$ only along some specific directions rather than all directions. These pseudo-extrema do not form isolated carrier pockets but will contribute to the electronic transport if they can strongly increase the anisotropy of the existing carrier pockets through connecting them. For $Mg_3Sb_2$, it is found that the valence band maximum (VBM) is located at the high-symmetry Γ point while the CBM occurs at the low-symmetry $CB_1$ point along the $M^*$-$L^*$ line inside the Brillouin zone (see Fig. 1 and Fig. 2).[24] One may easily overlook the accurate CBM by taking the U point along the M-L line as the location of the CBM since the band structure is normally plotted along high-symmetry lines. In reality, the U point is a pseudo-minimum point, where it is a minimum along the $a^*$ and $c^*$ directions but not a minimum along the $b^*$ direction (see Fig. 2a). The coordinate of the $CB_1$ point is (0, 0.417, 0.333), which might show a small deviation depending on the computational accuracy. According to the point group symmetry, there are six symmetry-equivalent positions for the $CB_1$ point, which leads to a high valley degeneracy of 6 (see Fig. 1).

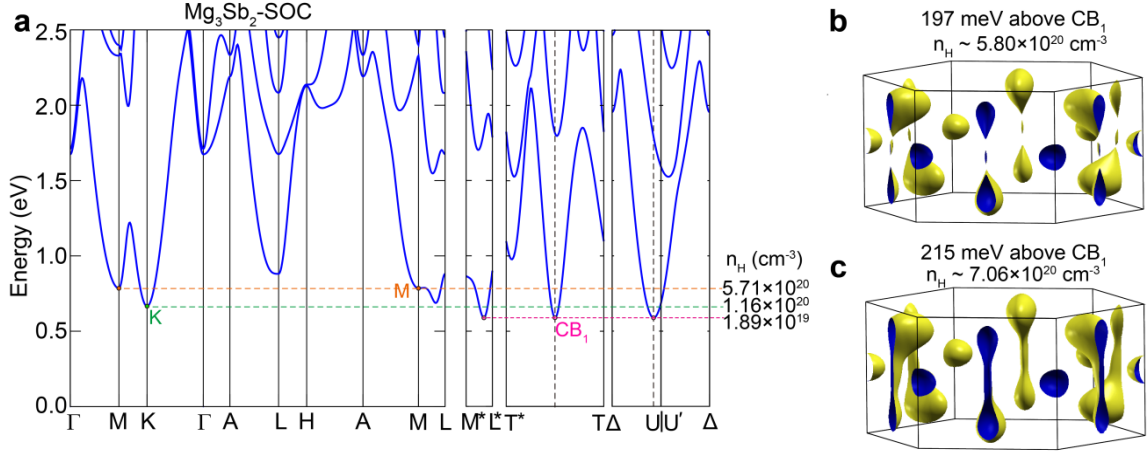

FIG. 2. (Color online) (a) Near-edge conduction bands of $Mg_3Sb_2$ calculated using the TB-mBJ potential including SOC in the Wien2k code. The band structure around the CBM at the $CB_1$ point is plotted along the $a^*$, $b^*$, and $c^*$ directions. The k paths along the $M^*$-$L^*$, $T^*$-T, $\Delta$-U|U'-$\Delta$ lines correspond to the $c^*$, $a^*$, and $b^*$ directions around the $CB_1$ minimum, respectively. (b) The iso-energy Fermi surface with a Fermi level 197 meV above the CBM. The small isolated pockets at the M point are highly elongated along the $c^*$ direction, indicating the high carrier pocket anisotropy. (c) The calculated Fermi surface with a Fermi level 215 meV above the CBM, which corresponds to the highest Fermi surface complexity factor $N_v^*K^*$ for n-type $Mg_3Sb_2$.

In addition to the CBM at the $CB_1$ point, there are another two near-edge conduction band minima at the K point ($N_v = 2$) and the M point ($N_v = 3$), respectively (see Fig. 2a). The band minimum at the K point is only about 76 meV above the CBM at the $CB_1$ point (see Table I), which is smaller than $\sim 2k_BT$ ($T \geq 450$ K). Considering the thermal broadening the Fermi distribution ($\sim 2k_BT$), the band minima $CB_1$ and K can be treated as effectively converged at elevated temperatures. Moving the Fermi level up to about 195 meV above the CBM, the band minimum at the M point is reached. The iso-energy Fermi surface corresponding to a Fermi level 2 meV above the band minimum M shows six half electron pockets at the M point, six full electron pockets at the $CB_1$, and six one-third pockets at the K point, which gives rise to a total valley degeneracy up to 11 (see Fig. 2b). However, the isolated electron pockets at the M point only exist within a very narrow energy window of $\sim 3$ meV above the M minimum. When the Fermi level moves up to slightly above 198 meV, the carrier pockets at the $CB_1$ and M point will connect and merge into large pockets along the $CB_1$-U-M-U' lines (see Fig. 2c). As a result, the anisotropy of carrier pockets is significantly increased, whereas the total effective degeneracy of conducting carrier pockets is decreased to 5. The corresponding Hall carrier concentrations at 300 K for the energy levels reaching the band minima $CB_1$, K, and M are $\sim 1.89 \times 10^{19}$, $\sim 1.16 \times 10^{20}$, and $\sim 5.71 \times 10^{20}$ cm$^{-3}$, respectively. Since the experimental Hall carrier concentration for n-type Te-doped $Mg_3Sb_2$ is about $0.21 \times 10^{19}$ cm$^{-3}$ at 300 K, the superior n-type electronic transport is dominantly contributed by the six-fold band minimum at the $CB_1$ point.

Using Boltzmann transport calculations based on the full *ab initio* band structures, the DOS effective mass and conductivity effective mass can be extracted from the theoretical Seebeck coefficient and electrical conductivity (see Section II). Unlike usually obtained by roughly fitting the near-edge electronic bands, the effective mass calculated here is averaged over all contributing electronic bands weighted by the Fermi distribution, which better characterizes the band structure by taking into account the nonparabolic and multiple band behaviors. As shown in Fig. 3a, the n-type doping shows larger $m_d^*$ values as well as smaller $m_c^*$ values at low doping concentrations and 300 K in comparison with those of the p-type doping in $Mg_3Sb_2$. This indicates the dominant contribution from the conduction band minimum $CB_1$ with a high valley degeneracy and a light conductivity effective mass, consistent with earlier reports.[18,19,24] Combining the calculated $m_d^*$ and $m_c^*$, we are able to quantitatively estimate the Fermi surface complexity factor $N_v^*K^* = (m_d^*/m_c^*)^{3/2}$ defined by Gibbs *et al.*,[52] where $N_v^*$ and $K^*$ are the effective valley degeneracy and effective anisotropy parameter of the carrier pockets, respectively. A higher value of $N_v^*K^*$ indicates a more complex Fermi surface and usually results in better electrical transport properties. Figure 3a shows the doping dependence of $N_v^*K^*$ at 300 K for p-type and n-type $Mg_3Sb_2$. A combination of the higher $m_d^*$ and lower $m_c^*$ values gives rise to remarkably larger $N_v^*K^*$ values with a maximum of ~19 for n-type $Mg_3Sb_2$, which is more than 4 times larger than that of p-type $Mg_3Sb_2$. Such a high value of $N_v^*K^*$ is comparable to those of the state-of-the-art TE materials such as $Bi_2Se_3$ and SnTe. As a result of the highly complex Fermi surface, the theoretical power factors for the n-type doping are strongly enhanced in comparison with those of the p-type doping in $Mg_3Sb_2$.

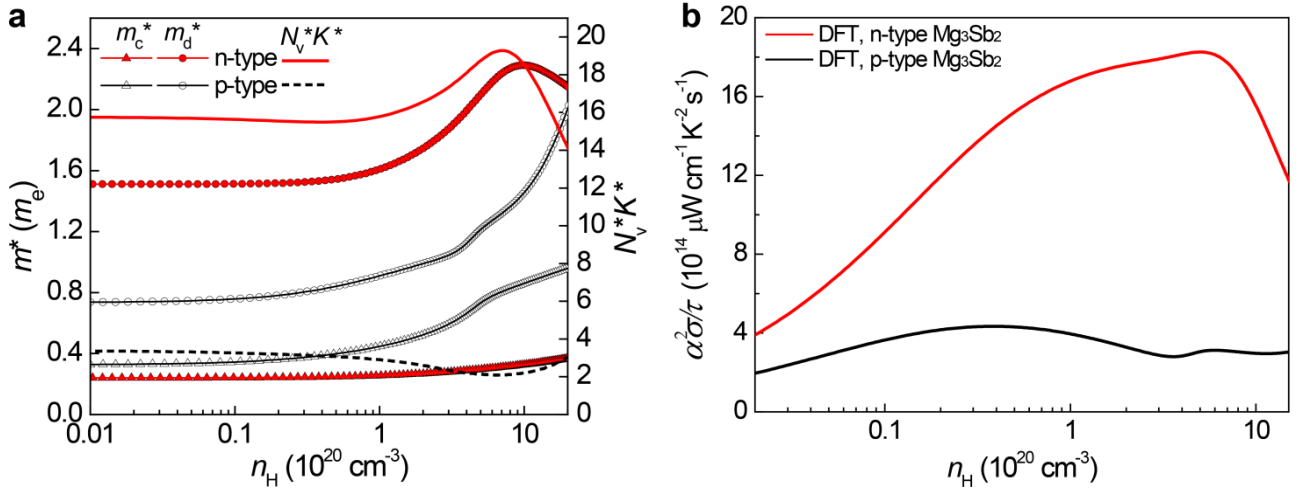

FIG. 3. (Color online) (a) Conductivity effective mass $m_c^*$, DOS effective mass $m_d^*$, and Fermi surface complexity factor $N_v^*K^*$ as a function of Hall carrier concentration ($n_H$) in p-type and n-type $Mg_3Sb_2$. (b) Theoretical power factor at 300 K as a function of Hall carrier concentration for p-type and n-type $Mg_3Sb_2$.[19]

In addition to the valley degeneracy, the anisotropy parameter of the carrier pocket $K^*$ is another origin for the complexity of Fermi surface. Highly anisotropic carrier pocket has been proved to result in potentially good electrical transport properties.[52-54] Anisotropic carrier pockets are commonly observed in many well-known TE materials. Even in simple cubic TE materials Si (Ge) and PbTe (or other IV-VI materials), we can see ellipsoidal carrier pockets for conduction band minima and valence band maxima, respectively. In n-type $Mg_3Sb_2$, the anisotropic feature of the $CB_1$ minimum is clearly seen from the ellipsoidal-like Fermi surface. To reveal the anisotropic feature, the near-edge band structures around $CB_1$ along the $a^*$ ($k_x$), $b^*$ ($k_y$), and $c^*$ ($k_z$) directions were calculated. The effective masses along the three directions for the $CB_1$ pocket are estimated by fitting the near-edge band dispersions. It should be noted that the effective masses obtained using this method are only used for the simplified discussion as it does not take into account the nonparabolic behavior. If we define the elongated direction of the carrier pocket as the longitudinal direction and the other two directions as the transverse directions, the effective masses for every individual carrier pocket at the $CB_1$ point can be represented as $m_\parallel^* = 0.55 m_e$, $m_{\perp,1}^* = 0.21 m_e$, and $m_{\perp,2}^* = 0.28 m_e$. For every single carrier pocket of $CB_1$, the effective mass is anisotropic along the three axial directions in the $k$ space, whereas after averaging over the six equivalent carrier pockets the average effective mass tensor is nearly isotropic and follows the crystal symmetry with $m_{kx}^* = m_{ky}^* = (0.21 \times 4 + 0.55 \times 2)/6 = 0.32 m_e$ and $m_{kz}^* = 0.28 m_e$. This is further confirmed by the result of the isotropic average effective mass tensor of n-type $Mg_3Sb_2$ at the low carrier concentration range ($n_H < 10^{20}$ cm$^{-3}$) extracted from the Boltzmann transport calculations (see Fig. S1). The nearly isotropic feature in the average effective mass tensor and n-type electrical transport properties can be attributed to the nearly isotropic orbital interactions, which are induced by the spherical $s$ orbitals of the Mg atoms that dominate the electronic states at the CBM. Despite the nearly isotropic feature in the average effective mass tensor, the benefit of the single carrier pocket anisotropy should not be overlooked. Taking the Fermi surface complexity factor $N_v^*K^*$ at 300 K and a low doping level of $1\times10^{18}$ cm$^{-3}$ where the CB$_1$ minimum plays a dominant role, the effective anisotropy factor of the carrier pockets can be estimated to be ~2.6, which make a moderate contribution to the superior $N_v^*K^*$ for n-type Mg$_3$Sb$_2$.

Table I. Lattice parameters, atomic coordinates, bandgaps ($E_g$), the energy difference between the near-edge conduction band minima of Mg$_3$Sb$_{2-x}$Bi$_x$ ($x$ = 0, 1, and 2). The lattice parameters $a$ and $c$ are experimental data from Refs. 25,55,56. Bandgaps and conduction band alignment are computed using the TB-mBJ potential with both the VASP and Wien2k codes for comparison. The SOC and noSOC denote the cases with and without considering the spin orbit coupling, respectively.

| Systems | $a$ (Å) | $c$ (Å) | $Z_{Mg(2)}$ | $Z_{Sb/Bi}$ | Codes | | $E_g$ (eV) | $\Delta E_{K\text{-}CB1}$ (eV) | $\Delta E_{M\text{-}CB1}$ (eV) | $\Delta E_{\Gamma\text{-}CB1}$ (eV) |
|---|---|---|---|---|---|---|---|---|---|---|
| Mg$_3$Sb$_2$ | 4.562 | 7.229 | 0.634 | 0.228 | VASP | noSOC | 0.63 | 0.112 | 0.222 | 1.120 |
| | | | | | | SOC | 0.50 | 0.111 | 0.223 | 1.120 |
| | | | | | Wien2k | noSOC | 0.70 | 0.075 | 0.194 | 1.084 |
| | | | | | | SOC | 0.59 | 0.076 | 0.195 | 1.084 |
| Mg$_3$SbBi | 4.608 | 7.315 | 0.626 | 0.227 | VASP | noSOC | 0.47 | 0.228 | 0.300 | 0.566 |
| | | | | | | SOC | 0.10 | 0.217 | 0.305 | 0.588 |
| | | | | | Wien2k | noSOC | 0.56 | 0.195 | 0.273 | 0.544 |
| | | | | | | SOC | 0.25 | 0.182 | 0.277 | 0.541 |
| Mg$_3$Bi$_2$ | 4.666 | 7.401 | 0.629 | 0.222 | VASP | noSOC | 0.31 | 0.352 | 0.377 | 0.092 |
| | | | | | | SOC | -0.20 | 0.340 | 0.390 | 0.092 |
| | | | | | Wien2k | noSOC | 0.41 | 0.317 | 0.352 | 0.060 |
| | | | | | | SOC | -0.09 | 0.312 | 0.362 | 0.060 |

N-type Mg$_3$Sb$_{2-x}$Bi$_x$ alloys are widely studied experimentally and theoretically owing to their outstanding TE performance. It is found that substituting Bi on the Sb sites do not destroy the multiple conduction band behavior since the near-edge conduction bands are dominated by the electronic states of the Mg atoms. However, increasing the Bi content in Mg$_3$Sb$_{2-x}$Bi$_x$ alloys will have a noticeable effect on the band gap and conduction band alignment. Table I shows the band gap and conduction band alignment of Mg$_3$Sb$_{2-x}$Bi$_x$ ($x$ = 0, 1, and 2). By comparing the results calculated by the VASP and Wien2k codes, it is found that the VASP code systematically predicts slightly smaller band gap values and larger energy differences between the near-edge conduction band minima. This is possibly induced by the VASP code using the pseudo density in calculating the $c$ parameter in the TB-mBJ method. In principle, the full-electron Wien2k code should give more accurate electronic structures than the VASP code. Therefore, the results calculated with the Wien2k code are used for the following discussions. Without considering the SOC, the band gap shows a gentle decrease with increasing the Bi content (see Table I and Fig. S2), which can be rationalized by an increase in band widths as the electronegativity difference between the Mg and the anion decreases using the molecular orbital scheme. When the SOC is included, the band gap undergoes a remarkable decrease from Mg$_3$Sb$_2$ (~0.59 eV) to Mg$_3$SbBi (~0.25 eV) to Mg$_3$Bi$_2$ (-0.09 eV, semimetal). The band-gap closure is estimated to occur at $x \approx 1.7$ for Mg$_3$Sb$_{2-x}$Bi$_x$. This is induced by the large spin orbit splitting of the Bi 6p states. Unlike the decreasing of the band gap in the cationic bismuth compound Bi$_2$S$_3$ that can be rationalized by the stabilization of Bi 6p$_{1/2}$ orbitals at the CBM,[57] the lowering band gap in Mg$_3$Sb$_{2-x}$Bi$_x$ ($x \neq 0$) alloys may be explained by the SOC destabilizing the Bi 6p$_{3/2}$ orbitals that contribute dominantly to the near-edge valence bands.

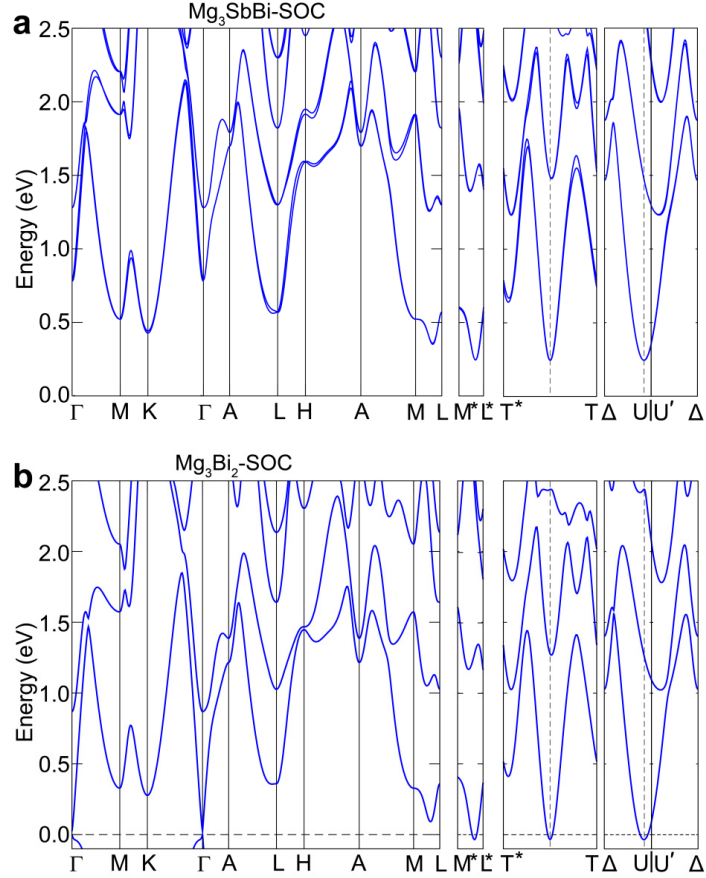

FIG. 4. (Color online) (a, b) Near-edge conduction bands of (a) Mg$_3$SbBi and Mg$_3$Bi$_2$ calculated using the TB-mBJ potential including SOC in the Wien2k code. The VBM is set to 0 eV.

Unlike the strong impact of the SOC on the band gaps, the SOC shows a negligible effect on the conduction band alignment in Mg$_3$Sb$_{2-x}$Bi$_x$ alloys (see Table I). Regarding the conduction band alignment, another band minimum at the Γ point ($N_v$ = 1) is taken into account as it shows a rapid downward shift approaching the CBM with increasing the Bi content (see Table 1 and Fig. 5(a)). However, the light band minimum Γ is about 0.2 eV above the CBM, which is too far from the CBM to play an important role in n-type electronic transport before the band-gap closure ($x \approx 1.7$). Even for Mg$_3$Bi$_2$, the band minimum Γ is still at about 0.06 eV above the CBM. Clearly, there is no band crossing between the band minima Γ and CB$_1$ in n-type Mg$_3$Sb$_{2-x}$Bi$_x$, which is inconsistent with the reports by Imasato et al.[25,29] The inconsistency is likely due to the earlier reports mistaking the pseudo-minimum at the U point as the CBM for discussions. For the alignment of the band minimum K, the Mg$_3$Bi$_2$ alloying pushes the band minimum K up to a bit far from the CBM, making the contribution of the band minimum K to near-edge n-type electronic transport insignificant. For the band minimum at the M point, increasing the Bi content also results in a larger energy difference between the band minima M and CB$_1$.

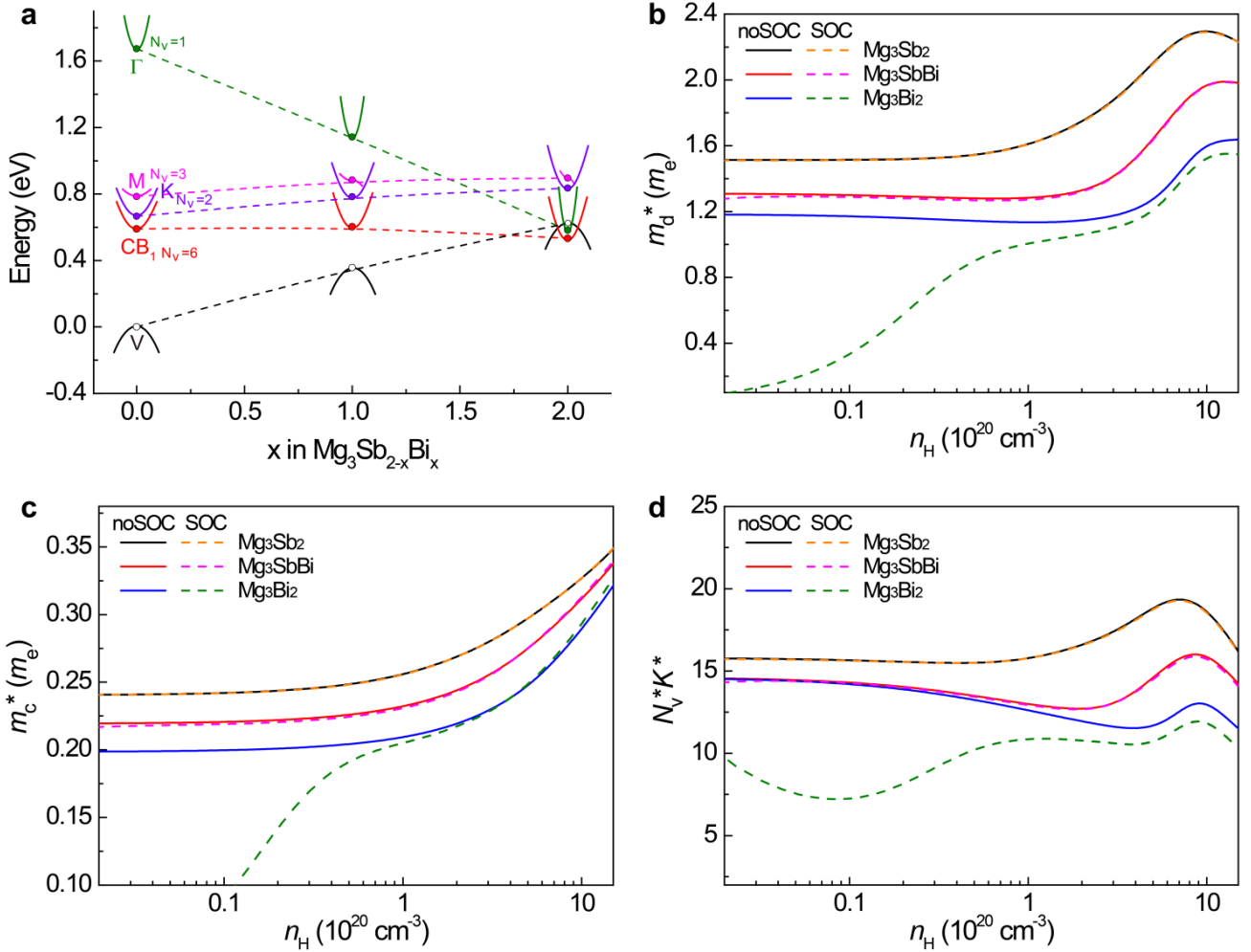

FIG. 5. (Color online) (a) Conduction band alignment of $Mg_3Sb_{2-x}Bi_x$ ($x = 0$, 1, and 2) calculated by the TB-mBJ potential considering the SOC in the Wien2k code. It should be noted that the four conduction band minima are located at different k points (i.e., $CB_1$, K, M, and Γ). Strictly speaking, the lowest conduction band at the M point is a band minimum in $Mg_3Sb_2$ but not in $Mg_3SbBi$ and $Mg_3Bi_2$. The band alignment is obtained by aligning the Mg-1s core levels of different compounds. The VBM of $Mg_3Sb_2$ at the Γ point is set to 0 eV. (b) Density of states effective mass, (c) conductivity effective mass, and (d) Fermi surface complexity factor $N_v^*K^*$ as a function of Hall carrier concentration at 300 K for n-type $Mg_3Sb_{2-x}Bi_x$. SOC and noSOC denote the cases with and without spin orbit coupling, respectively.

Moreover, increasing the $Mg_3Bi_2$ content in $Mg_3Sb_{2-x}Bi_x$ alloys gives rise to a moderate change in the conduction band curvature. As can be seen in Fig. 4 and Fig. S2, the significant decrease of the band gap with increasing the Bi composition is accompanied by the increasing near-edge conduction band widths and more dispersive CBM. As a result, the effective masses $m_d^*$ and $m_c^*$ at 300 K show a decreasing trend from $Mg_3Sb_2$ to $Mg_3SbBi$ to $Mg_3Bi_2$ (see Figs. 5(b) and 5(c)). This is consistent with the earlier reports.[25,37] Since smaller effective valley degeneracy induced by larger energy separation between near-edge band minima also contributes to the reduction in $m_d^*$ (especially at elevated temperatures), $m_d^*$ typically shows a larger decrease than $m_c^*$ with increasing the Bi content. Thus, the Fermi surface complexity factor $N_v^*K^*$ shows a decreasing trend with increasing the Bi content (see Fig. 5(d)). Without considering the SOC, within the low carrier concentration range the $N_v^*K^*$ values in $Mg_3Bi_2$ are comparable with those in $Mg_3SbBi$, which can be attributed to the moderate contribution of the Γ band minimum in $Mg_3Bi_2$ due to the small energy difference between the band minima Γ and $CB_1$. However, considering the effect of the SOC results in noticeably decreasing $m_d^*$, $m_c^*$, and $N_v^*K^*$ at $n_H < 10^{20}$ cm$^{-3}$ in $Mg_3Bi_2$, which is typically induced by the strong bipolar effect. The effective masses $m_d^*$ and $m_c^*$ are nearly independent of carrier concentration at the low doping concentration region and then start to increase at specific doping concentrations. The onset of increasing effective masses occurs at a higher carrier concentration for the compound with a higher Bi content, which can be understood by a larger energy difference between the band minimum K and $CB_1$. As confirmed in the temperature-dependent DOS effective mass of $Mg_3Sb_2$ (see Fig. S3), the contribution of the band minimum K can be very significant within low carrier concentrations at elevated temperatures due to the increasing thermal broadening of Fermi distribution.

Interestingly, $m_d^*$ and $N_v^*K^*$ of n-type $Mg_3Sb_{2-x}Bi_x$ show characteristic peaks within the Hall carrier concentration range of ~0.5-1.5×10$^{21}$ cm$^{-3}$ and the corresponding energy levels of these optimal carrier concentrations are close to the band minimum M (see Figs. 5(b) and 5(d)). The optimal energy levels for the peaks of $N_v^*K^*$ are slightly above the band minimum M. The corresponding iso-energy Fermi surfaces show the $CB_1$ pockets being connected along the $CB_1$-U-M-U′ lines (see Fig. 2(c)), which considerably increases the carrier pocket anisotropy and thereby gives rise to the peak in $N_v^*K^*$. It should be noted that the peak value of $N_v^*K^*$ shows a remarkable decrease from $Mg_3Sb_2$ to $Mg_3SbBi$ to $Mg_3Bi_2$. This may be understood from the fact that the length of the thread connecting the $CB_1$ pockets gets smaller with increasing the Bi content as the conduction band along M-L becomes more dispersive (see Figs. 2(a) and 4), which will result in decreasing carrier pocket anisotropy and $N_v^*K^*$. Moreover, there is another favorable band feature that results in a superior optimal value of $N_v^*K^*$ in n-type $Mg_3Sb_2$. As mentioned previously, for $Mg_3Sb_2$ the energy of the lowest conduction band increases by about 3 meV and then decreases along the M-L line, which indicates a band minimum at the M point. This leads to the formation of the isolated pockets at the M point for an energy level within ~3 meV above the band minimum M (see Fig. 2(b)), which maximizes the valley degeneracy up to 11 and effectively improves $N_v^*K^*$. However, for $Mg_3SbBi$ and $Mg_3Bi_2$ there is no formation of the isolated pockets at the M point since the lowest conduction bands at the M point is not a band minimum but a band maximum along the M-L line.

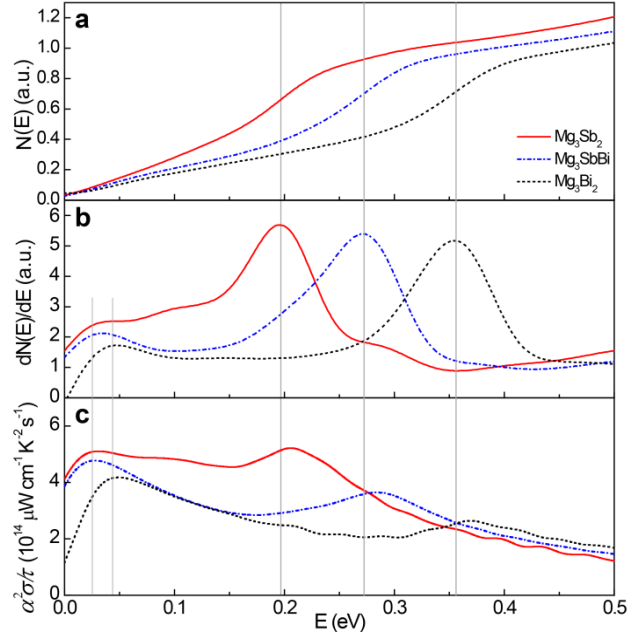

FIG. 6. (Color online) (a) Density of states $N(E)$, (b) the energy derivative of the density of states $dN(E)/dE$, and (c) the energy-dependence of theoretical power factors at 150 K of $Mg_3Sb_2$, $Mg_3SbBi$, and $Mg_3Bi_2$, which are based on the electronic structure calculations with the TB-mBJ potential considering the SOC in the Wien2k code. In order to reduce the effect of the thermal broadening of the Fermi distribution, here the theoretical power factor at a low temperature of 150 K is chosen for comparison with the DOS. The CBM at the $CB_1$ point is set to 0 eV.

Carrier pockets connected by threads have been proved to result in a steeply increasing density of states, enhanced Seebeck coefficient, and the superior electronic transport in PbTe.[58] To gain additional insight into the effect of the connected pockets on the DOS and electrical transport properties, we calculate and compare the DOS of the near-edge conduction bands of $Mg_3Sb_2$, $Mg_3SbBi$, and $Mg_3Bi_2$ (see Fig. 6(a)). The decreasing DOS near the CBM is a further confirmation of the decreasing effective mass of the near-edge conduction bands with increasing the Bi composition. The relation between the DOS and Seebeck coefficient can be described by the Mott expression:[59]

$$\alpha = \frac{\pi^2 k_B^2 T}{3e} \left\{ \frac{1}{n} \frac{dn(E)}{dE} + \frac{1}{\mu} \frac{d\mu(E)}{E} \right\}_{E=E_F}, \quad (2)$$

where $\mu(E)$ is the mobility at the energy level $E$, and carrier density $n(E)$ can be written as $N(E)f(E)$ with $N(E)$ and $f(E)$ being, respectively, the electronic density of states and Fermi function at the energy level $E$. The Mott equation shows one effective way to improve the Seebeck coefficient intrinsically from band structure: a local increase in the density of states $N(E)$ around the Fermi level $E_F$.

Figure 6(b) shows the energy derivative of the DOS of the conduction bands for $Mg_3Sb_{2-x}Bi_x$ ($x$ = 0, 1, and 2). As expected, the maximum of d$N(E)$/d$E$ (in other words, the steepest enhancement of the DOS) occurs at the energy level very close to the band minimum M, where the flat conduction bands close to the M point along the M-L line will form threads connecting the $CB_1$ carrier pockets. The steep increase of the DOS induced by the connection of the $CB_1$ carrier pockets is clearly conducive to obtaining enhanced Seebeck coefficient and power factor. This is confirmed by the result of the theoretical power factor $\sigma/\tau$ at 150 K that shows a characteristic peak at an energy level slightly above the band minimum M (see Fig. 6(c)), which corresponds well with the optimal energy level that the peak of $N_v^* K^*$ occurs. As the Bi content increases in $Mg_3Sb_{2-x}Bi_x$, the peak related to the band minimum M is shifted to a higher energy level and the peak value of the theoretical power factor decreases. This can be understood from a larger energy separation between band minima M and $CB_1$ as well as smaller peak values in $N_v^* K^*$ for the compound with a larger Bi content. In addition to the peak around the band minimum M, there is another peak located at a low energy level slightly above the $CB_1$. This peak is likely contributed by the six-fold $CB_1$ minimum with the light conductivity effective mass. For $Mg_3Sb_2$ the optimal power factor occurs at the maximum of the peak at ~0.20 eV close to the band minimum M, whereas for $Mg_3SbBi$ and $Mg_3Bi_2$ the optimal power factor occurs at the peak close to the band minimum $CB_1$.

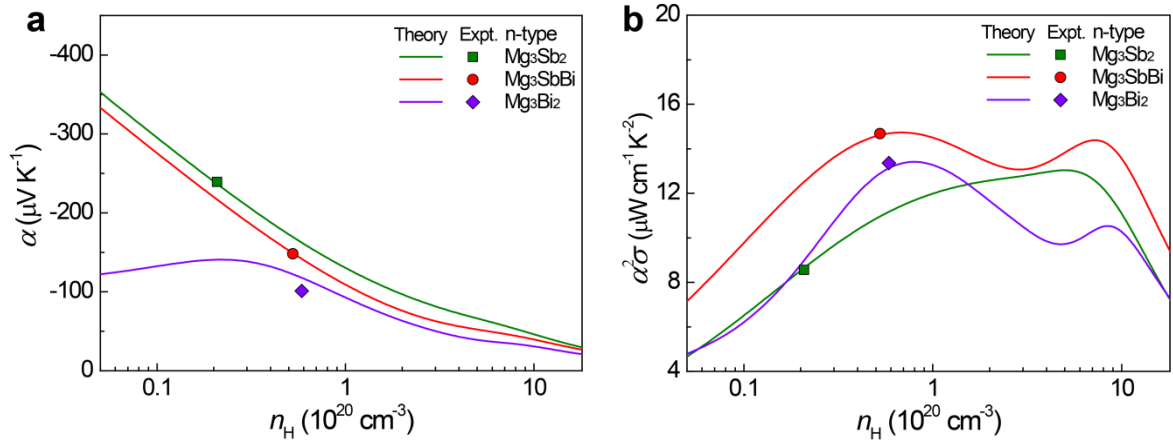

FIG. 7. (Color online) (a) Seebeck coefficient and (b) power factor at 300 K as a functional Hall carrier concentration for n-type $Mg_3Sb_{2-x}Bi_x$ ($x$ = 0, 1, and 2). The theoretical electrical transport data are based on the full *ab initio* electronic structure calculations using the TB-mBJ potential considering the SOC in the Wien2k code. The experimental data of n-type $Mg_3Sb_{2-x}Bi_x$ ($x$ = 0, 1, and 2) are taken from Ref. 29. The carrier scattering time $\tau$ is estimated by comparing the experimental electrical conductivity $\sigma$ with the theoretical $\sigma/\tau$.

Figure 7 shows the Seebeck coefficients and power factors as a function of Hall carrier concentration at 300 K

for n-type $Mg_3Sb_2$, $Mg_3SbBi$, and $Mg_3Bi_2$. It is clear that increasing the Bi content in n-type $Mg_3Sb_{2-x}Bi_x$ lowers the Seebeck coefficient through reducing the DOS effective mass, which is mainly caused by reducing the conduction band mass and weakening the contribution of the secondary band minimum K. For the semimetal $Mg_3Bi_2$, the bipolar effect is also found to contribute to the rapid reduction of the Seebeck coefficient especially when the carrier concentration is lower than $\sim 3\times 10^{19}$ cm$^{-3}$. However, reducing the conductivity effective mass is beneficial to enhancing the carrier mobility, weighted mobility, and eventually power factor when the bipolar effect is not significant. This is confirmed by both the experimental and theoretical results that show higher power factors in $Mg_3SbBi$ in comparison with those of $Mg_3Sb_2$. In addition, for the n-type doping in $Mg_3Sb_2$, the experimental carrier concentration of $\sim 2\times 10^{19}$ cm$^{-3}$ is still far below the theoretical optimal value of $\sim 5\times 10^{20}$ cm$^{-3}$. The power factor of n-doped $Mg_3Sb_2$ can be further improved and optimized by tuning the carrier density to the optimal value, at which the carrier pocket anisotropy and Fermi surface complexity $N_v^*K^*$ are strongly enhanced. However, for $Mg_3SbBi$ and $Mg_3Bi_2$ the experimental carrier concentrations are already very close to the optimal values. This is likely an additional benefit for the $Mg_3Bi_2$ alloying to result in the superior electrical transport performance at low temperatures.

## IV. CONCLUSION

In this work, we have investigated the conduction band alignment, the effective mass, and the Fermi surface complexity, and their relationship with electronic transport in n-type $Mg_3Sb_{2-x}Bi_x$ ($x$ = 0, 1, and 2) based on full *ab initio* band structure calculations. According to the result of the conduction band alignment, we find that the CBM is dominated by the favorable six-fold band minimum $CB_1$. Although the Γ band minimum undergoes a rapid downward shift with increasing the Bi content, it is still too far from the CBM to play an important role in near-edge n-type electronic transport before the band-gap closure. There is no band crossing between the $CB_1$ and Γ band minima in n-type $Mg_3Sb_{2-x}Bi_x$. In contrast to the Γ band minimum approaching the CBM, another two conduction band minima at the K and M points both show larger energy separations with the $CB_1$ minimum with increasing the Bi content

The conductivity effective mass and density of states effective mass have been extracted from the Boltzmann transport calculations. The Fermi surface complexity factor $N_v^*K^*$, a combination of the effective valley degeneracy $N_v^*$ and carrier pocket anisotropy $K^*$, has been evaluated from the ratio the two types of effective masses. The anisotropy of the dominant six-fold $CB_1$ pockets is found to contribute to the superior Fermi surface complexity factor at the low doping level. The connection of the six-fold carrier pockets $CB_1$ at the energy level reaching the band minimum M is able to strongly enhance the carrier pocket anisotropy and Fermi surface complexity, which is likely responsible for the local peak in the theoretical power factor. Accordingly, for $Mg_3Sb_2$ the power factor can be further optimized if the doping level is able to approach the band minimum M. Moreover, we have studied the influence of the increasing Bi content on the electronic structure parameters and electrical transport properties. It is found that increasing the Bi composition reduces the density of states effective mass, Seebeck coefficient, and Fermi surface complexity factor. The narrowing of the band gap with increasing the Bi content also indicates an increasing detrimental effect from the bipolar conduction for n-type $Mg_3Sb_{2-x}Bi_x$ ($x$ > 1). However, increasing the Bi content in n-type $Mg_3Sb_{2-x}Bi_x$ indeed gives rise the reduction of the conductivity effective mass, which is conducive to improving the power factor if there is no noticeable bipolar effect.

## SUPPLEMENTARY MATERIAL

See the supplementary material for the average conductivity effective mass tensor at 300K for n-type $Mg_3Sb_2$,

the near-edge conduction bands without considering the spin orbit coupling for all compounds, the density of states effective mass at various temperatures for n-type $Mg_3Sb_2$, and the comparison of structural parameters, band gaps, and conduction band alignments using different structural optimization methods.

## ACKNOWLEDGEMENTS

This work was supported by the Danish National Research Foundation (Center for Materials Crystallography, DNRF93) and the Danish Center for Scientific Computing. The numerical results presented in this work were obtained at the Center for Scientific Computing, Aarhus.